\documentclass[10pt,conference]{IEEEtran}

\ifCLASSOPTIONcompsoc
  \usepackage[nocompress]{cite}
\else
  \usepackage{cite}
\fi
\ifCLASSINFOpdf
\else
\fi

\usepackage[table, svgnames]{xcolor}

\usepackage{url}
\usepackage{array}
\usepackage{xcolor}
\usepackage{cancel}
\usepackage{siunitx}
\usepackage{multirow}
\usepackage{graphicx}
\usepackage{textcomp}
\usepackage{multirow}
\usepackage{stfloats}
\usepackage{algorithm}
\usepackage[T1]{fontenc}
\usepackage[utf8]{inputenc}
\usepackage{mathtools, cuted}
\usepackage{amsmath,amsfonts}
\usepackage[noend]{algpseudocode}
\usepackage{soulutf8}
\usepackage{helvet}
\usepackage{multicol}
\usepackage{graphicx}
\usepackage{subcaption}

\DeclareRobustCommand*{\IEEEauthorrefmark}[1]{%
  \raisebox{0pt}[0pt][0pt]{\textsuperscript{\footnotesize #1}}%
}

\IEEEoverridecommandlockouts

\graphicspath{{./figs/}}

\newlength{\textfloatsepsave} 
\begin{document}

\title{RESAA: A Removal and Structural Analysis Attack Against Compound Logic Locking}

        \author{
		\IEEEauthorblockN{Felipe~Almeida\IEEEauthorrefmark{1}, Levent~Aksoy\IEEEauthorrefmark{1} and Samuel~Pagliarini\IEEEauthorrefmark{1}\,\IEEEauthorrefmark{2}}
		\IEEEauthorblockA{
            \IEEEauthorrefmark{1}Department of Computer Systems, Tallinn University of Technology, Tallinn, Estonia\\
            \IEEEauthorrefmark{2}ECE Department, Carnegie Mellon University, Pittsburgh - PA, USA\\
            }
		Email: \IEEEauthorrefmark{1}{\{felipe.almeida, levent.aksoy, samuel.pagliarini\}}@taltech.ee \\
            Email: \IEEEauthorrefmark{2}{pagliarini}@cmu.edu
		\vspace{-4mm}
	}

    \maketitle
    
    \begin{abstract}
    The semiconductor industry's paradigm shift towards fabless integrated circuit (IC) manufacturing has introduced security threats, including piracy, counterfeiting, hardware Trojans, and overproduction. In response to these challenges, various countermeasures, including Logic locking (LL), have been proposed to protect designs and mitigate security risks. LL is likely the most researched form of intellectual property (IP) protection for ICs. A significant advance has been made with the introduction of compound logic locking (CLL), where two LL techniques are concurrently utilized for improved resiliency against attacks. However, the vulnerabilities of LL techniques, particularly CLL, need to be explored further. This paper presents a novel framework, RESAA, designed to classify CLL-locked designs, identify critical gates, and execute various attacks to uncover secret keys. RESAA is agnostic to specific LL techniques, offering comprehensive insights into CLL's security scenarios. Experimental results demonstrate RESAA's efficacy in identifying critical gates, distinguishing segments corresponding to different LL techniques, and determining associated keys based on different threat models. In particular, for the oracle-less threat model, RESAA can achieve up to 92.6\% accuracy on a relatively complex ITC'99 benchmark circuit. The results reported in this paper emphasize the significance of evaluation and thoughtful selection of LL techniques, as all studied CLL variants demonstrated vulnerability to our framework. RESAA is also open-sourced for the community at large. 

    \end{abstract}

    

	\section{Introduction}
\label{sec:introduction}

The evolution of the semiconductor industry and the migration to a fabless integrated circuit (IC) ecosystem has been revolutionary~\cite{shaw2000}. Outsourcing IC fabrication to third-party foundries and incorporating third-party intellectual properties (IPs) has significantly transformed the security dynamics in chip design. This shift has brought forth a spectrum of security threats, such as IC counterfeiting, IP piracy, IC overproduction, and the insertion of hardware Trojans. All these threats undermine the integrity of the IC supply chain~\cite{rostami2014}.

Counterfeiting leads to the unauthorized replication of ICs, which can be associated with a loss in both quality and reliability~\cite{zarrinchian2023}. Piracy involves the illicit use of IP fueling the production of counterfeit ICs~\cite{rostami2014}. Overproduction increases the proliferation of counterfeit products by manufacturing beyond authorized quantities~\cite{guin2014}. Hardware Trojans are pieces of malicious logic inserted into a design, potentially compromising its functionality and/or reliability~\cite{supon2021}.

\begin{figure}[t]
	\centerline{\includegraphics[width=9.5cm]{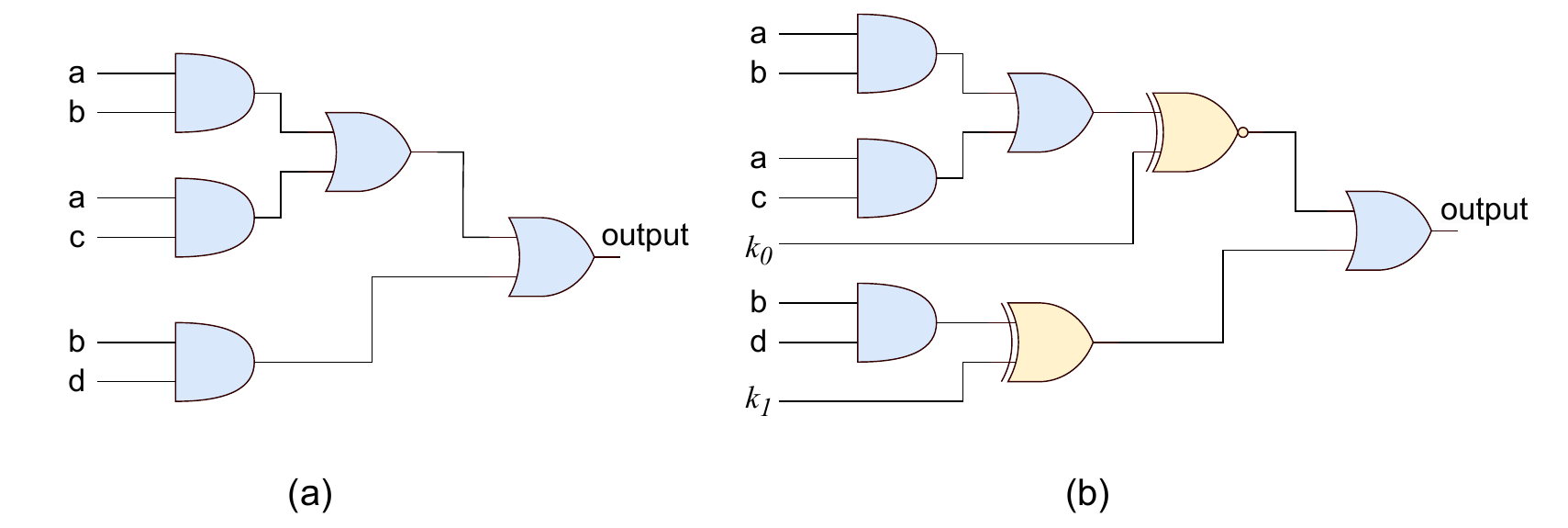}}
        \caption{(a) Original circuit; (b) Locked circuit where the secret key is $k_0k_1 = 10$.}
 \vspace{-4mm}
	\label{fig:logiclocking}
\end{figure}

As chip design and fabrication grows increasingly complex, maintaining the integrity of ICs and their IPs has emerged as a preeminent concern. In response to these emerging threats, researchers have proposed diverse countermeasures, such as split manufacturing, hardware metering, watermarking, and logic locking (LL). In split manufacturing design, the metal stack is divided across different foundries to mitigate security risks~\cite{perez2020}. Hardware metering involves real-time monitoring of resource usage within IC against piracy using mechanisms designed to track and regulate the allocation of hardware resources, ensuring efficient utilization while maintaining security~\cite{koushanfar2001}. Watermarking embeds signatures into designs, without altering functionality, to detect IP theft and misuse~\cite{qu2012}. LL stands out as likely the most researched technique; however, it offers only \emph{potential} protection against several security threats~\cite{roy2008, yasin2016, nguyen2021, dupuis2019, azar2019}. The principle behind LL is the insertion of additional logic driven by key bits, such that the locked circuit behaves like the original circuit only when the secret key is provided, as shown in Fig.~\ref{fig:logiclocking}.

Over nearly two decades, researchers have strived to improve LL techniques by developing strategies that promote output corruption, deliver resilience against attacks, and reduce associated overheads~\cite{kamali2022}. Specific LL methodologies, like random logic locking (RLL), have been fine-tuned to simultaneously reduce area overhead and increase output corruption. Fig.~\ref{fig:logiclocking} shows an example of RLL, which involves the insertion of additional gates controlled by key inputs, ensuring proper functionality solely when the secret key is applied~\cite{roy2008}. However, its security has been compromised, primarily by attacks under the oracle-guided (OG) threat model~\cite{subramanyan2015}. The most prominent of these attacks is known as the SAT-based attack~\cite{subramanyan2015}. 

\begin{figure}[t]
	\centerline{\includegraphics[width=1\linewidth]{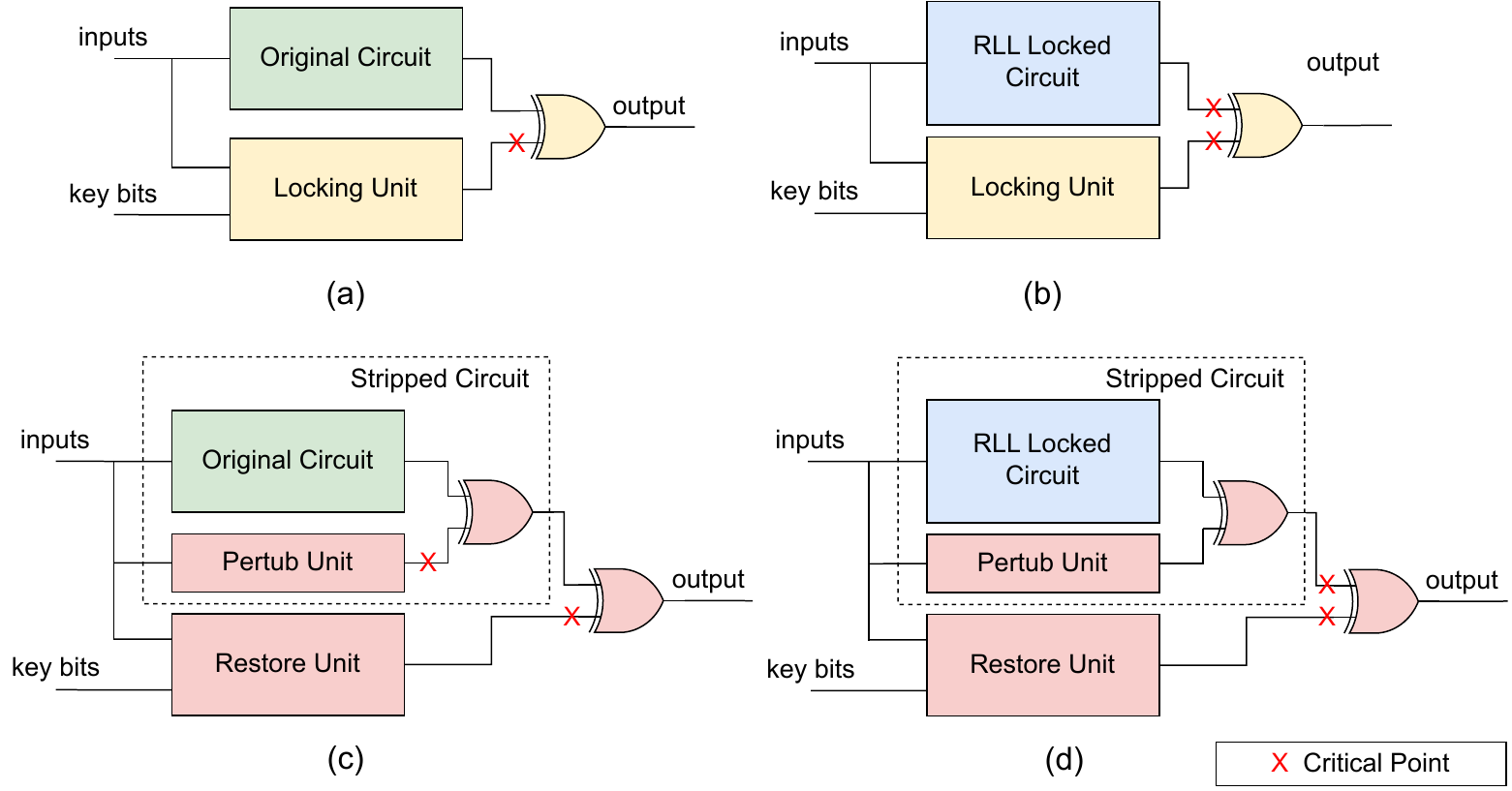}}
	\caption{High-level architecture of (a) SFLT (b) RLL +  SFLT (c) DFLT, and (d) RLL + DFLT in a CLL scheme. The critical signals are indicated by ``X'' in red.}
	\vspace{-4mm}
	\label{fig:circuits}
\end{figure}

To counter the SAT attack, SAT-resilient techniques have been developed under the name of provably secure logic locking (PSLL)~\cite{yasin2017_2}. A single-flip locking technique (SFLT) is one that incorporates a point function that introduces an additional block (locking unit), activated by a key input, to secure a specific output of the IC~\cite{xie2019,shakya2019,yasin2016}. Conversely, a double-flip locking technique (DFLT) also employs a point function but enhances security by inserting a perturb unit and a restore unit, followed by output correction~\cite{yasin2017,sengupta2020_0}, as shown in Fig.~\ref{fig:circuits} (a) and Fig.~\ref{fig:circuits} (c), respectively. Although these techniques aim to bolster resilience against SAT specifically, they remain vulnerable to various attacks, under both OG and oracle-less (OL) threat models. Additionally, certain variants may be susceptible to removal and structural analysis attacks.

Approaches to address the perceived weakness of LL techniques have taken many forms, including the deliberate insertion of cyclic logic~\cite{rezaei2018}, the use of emerging materials~\cite{divyanshu2022}, and look-up table based obfuscation~\cite{kamali2018,abideen2024}. A hybrid approach has emerged, too, termed compound logic locking (CLL), which combines high-output corruptibility and SAT-resilient LL techniques to address vulnerabilities that are present in obfuscated circuits when either technique is utilized alone~\cite{john2020,limaye2021}. As a general trend, there is limited knowledge about attacks that are CLL-aware. In~\cite{john2020}, authors have explored the combined use of SAT-based and structural attacks against CLL. It is worth noting that current CLL research is limited to a single combination of techniques, which highlights the need for a more comprehensive exploration. 

\subsection{Scope of this work}

This work explores attacks on CLL, an advanced IP protection approach characterized by a multi-layered application of existing LL techniques. We examine the effects of combining two LL techniques to enhance the security of digital circuits. In this scenario, we first lock the original circuit using RLL and then subsequently apply a PSLL technique. Although CLL is generally described as a more resilient approach, our study reveals that it may not necessarily mitigate known vulnerabilities. Instead, CLL could perpetuate or even worsen weaknesses inherent to the individual LL techniques. In some cases, combining SAT-resilient techniques with traditional LL has still left designs vulnerable to removal attacks, where attackers can bypass the locking mechanisms without solving the key. We aim to clarify how, in specific scenarios, CLL may fail to address these vulnerabilities, potentially fully exposing the original circuit.

We put forward RESAA, a comprehensive framework for: (i) identifying critical gates (CGs), (ii) classifying locked designs based on LL techniques, (iii) partitioning the design to apply well-established attacks, and (iv) potentially exposing secret keys using the identified LL technique. Our investigations reveal potential security vulnerabilities inherent in CLL compared to using a singular LL technique. These findings emphasize the intricate complexities and associated pitfalls of CLL strategies, highlighting the importance of thorough evaluation and careful use of LL techniques.

In contrast to previous studies~\cite{john2020,limaye2021}, our methodology takes a practical approach by relying solely on commercial synthesis tools. In other words, RESAA is a framework for security analysis that can be readily used in existing design flows. Furthermore, all the circuits we study in this work are mapped to a commercial cell library for the sake of realism. This industry-minded approach to our CLL analysis provides insights that are directly applicable to real-world situations, ensuring that our findings address the challenges and limitations encountered by design houses attempting to protect their commercial IP. 

\subsection{Contributions}

\begin{enumerate}

\item We present the RESAA framework, which identifies CGs in CLL designs and differentiates between RLL and PSLL techniques.

\item RESAA effectively divides designs into two parts, enabling the use of OL and OG attacks to uncover the secret key, exposing vulnerabilities in multiple CLL techniques.

\item RESAA is integrated within an industry-grade logic synthesis tool to ensure a realistic setting.

\item We open source scripts and strategies utilized in RESAA\footnote{See \url{https://github.com/Centre-for-Hardware-Security/CLL_attack}}. 

\item We expose the keys for large CLL-locked combinational circuits, even under the difficult OL setting, with a high accuracy.


\end{enumerate}

It is also important to emphasize that our approach does not assume that the adversary is aware of the combination of techniques that make up the chosen CLL scheme. Instead, we analyze and classify netlists in order to automatically identify the SFLT/DFLT technique being employed. This aspect significantly \emph{distinguishes} our work from existing methodologies that rely on assumptions.


The subsequent sections of this paper are structured as follows: Section~\ref{sec:background} elaborates more on the foundational concepts surrounding (C)LL. The RESAA framework is described in full detail in Section~\ref{sec:methodology}. Our experimental results are thoroughly covered in Section~\ref{sec:results}. Finally, Section~\ref{sec:conclusions} summarizes our findings.

\begin{figure*}[t]
	\centerline{\includegraphics[width=18.0cm]{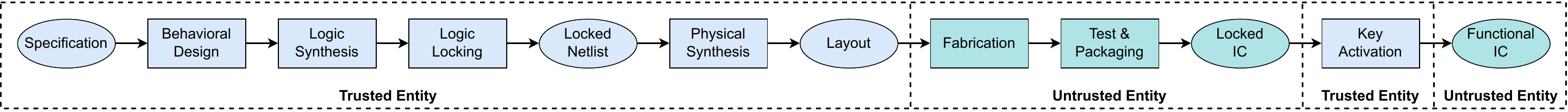}}
	\caption{Conventional logic locking in the IC design flow (adapted from~\cite{yasin2017_2}).}
 \vspace{-2mm}
	\label{fig:icflow}
\end{figure*}

	\section{Background and Related Work}
\label{sec:background}

This section presents details on LL, the OG and OL threat models, existing defenses and attacks, and the never-ending cat-and-mouse game between attackers and defenders.

\subsection{Logic Locking and Threat Models}

Fig.~\ref{fig:icflow} presents the conventional IC design flow augmented by using LL. The locking process is a security measure typically implemented at the gate level. LL aims to encode the circuit's functionality using key inputs, ensuring that unauthorized parties cannot access or replicate the design's core functionality without the corresponding secret key. Notably, implementing LL allows the design house to protect its proprietary information while engaging with third-party foundries for manufacturing. Once the locked IC design is finalized, the layout is sent to the foundry for fabrication. Upon completion of manufacturing, the values of the secret key are securely stored within a memory that is designed to resist unauthorized access and tampering attempts. Subsequently, once the functional ICs are prepared for market distribution, the locked design and its associated secret keys remain securely protected within the confines of the design house, thereby mitigating the risks associated with IP piracy and unauthorized overproduction.

LL is a countermeasure extensively researched to protect IP against several threats in the semiconductor industry. This technique involves strategically inserting extra logic gates into the design~\cite{roy2008, yasin2016, nguyen2021, dupuis2019, azar2019}. These added gates operate under the control of key inputs, ensuring that the locked design behaves the same as the original one if the secret key is applied. Otherwise, it generates a wrong (corrupted) output. LL has displayed both effectiveness and vulnerability against various attack methodologies, prompting the development of dozens of variants of the technique~\cite{mellor2021}.

Threat models play a critical role in identifying potential adversaries and their capabilities, which is essential for securing hardware designs against attacks. The existing literature considers two distinct threat models: the OG and OL threat models. Under both threat models, the adversary is assumed to have the locked netlist. Moreover, under the OG threat model, the adversary has a functional IC, which is used as an oracle to apply inputs and observe outputs. However, the oracle does not expose the key directly.

In this work, we exercise adversarial capabilities in \textbf{both OG and OL threat models}. We assume an adversary with access to a modern logic synthesis tool. We also assume the adversary can trivially differentiate key inputs from primary inputs\footnote{This is considered possible by tracing the key inputs back to the tamper-proof memory.}. However, the adversary does not know \emph{a priori} which key input corresponds to which LL technique in CLL. Finally, we assume the adversary is familiar with existing attacks and can apply them freely, including the following attacks: SAT-based~\cite{subramanyan2015}, AppSAT~\cite{shamsi2017}, SCOPE~\cite{scope2020}, DoubleDIP~\cite{sat2018}, and query-based~\cite{qatt2023} methods. The attack itself can be carried out on a standard configuration laptop or desktop, with no need for specialized computational resources or servers. Additionally, the adversary is assumed to have access to the standard cell library if resynthesis is sought\footnote{This assumption is reasonable if the adversary is located at the foundry.}.

Furthermore, a more nuanced approach to determining adversarial knowledge is the classification of specific adversaries (SA), knowledgeable adversaries (KA), and oblivious adversaries (OA). As the names imply, there are varying levels of awareness regarding the LL techniques employed. Table~\ref{tab:respectives} compares various tools/attacks for exploring vulnerable LL designs. The first tools were developed for specific LL techniques, with the adversary designing the attack exclusively for those techniques (therefore classified as SA). For example, SAT-based attacks~\cite{subramanyan2015} and AppSAT~\cite{shamsi2017} were initially aimed at RLL, while Double DIP specifically targeted SARLock designs. However, these attacks have been adapted and can now be applied to a broader range of LL techniques. Fa-SAT~\cite{limaye2021} was one of the first studies to explore CLL attacks, but only for a few specific combinations of LL techniques, such as bilateral logic encryption (BLE)~\cite{rezaei2020} and Anti-SAT coupled with RLL. In contrast, Valkyrie~\cite{limaye2022} proposed a framework to attack fifteen LL techniques, which yielded satisfactory results, but the adversary knew the techniques employed in advance (therefore classified as a KA tool). Our study explores a more generic tool capable of achieving remarkable results without \emph{a priori} knowledge about the CLL techniques employed (therefore, RESAA is classified as OA).

\begin{table}[t]
  \centering
  \footnotesize
  \caption{SA, KA, and OA attacks.}
  \vspace{-2mm}
  \begin{tabular}{|@{\hskip3pt}c@{\hskip3pt}|c@{\hskip3pt}|c@{\hskip3pt}|c@{\hskip3pt}|}
    \hline
    Attack & SA & KA &  OA \\ 
\hline
SAT-based & \textcolor{green}{\checkmark} &  \textcolor{red}{\texttimes} &  \textcolor{red}{\texttimes}\\
AppSAT & \textcolor{green}{\checkmark} &  \textcolor{red}{\texttimes} &  \textcolor{red}{\texttimes}\\
Double DIP& \textcolor{green}{\checkmark} &  \textcolor{red}{\texttimes} &  \textcolor{red}{\texttimes}\\
Fa-SAT & \textcolor{green}{\checkmark} &  \textcolor{red}{\texttimes} &  \textcolor{red}{\texttimes}\\
Valkyrie & \textcolor{green}{\checkmark} &  \textcolor{green}{\checkmark} &  \textcolor{red}{\texttimes}\\
This Work (RESAA) & \textcolor{green}{\checkmark} &  \textcolor{green}{\checkmark} &  \textcolor{green}{\checkmark}\\
\hline
\end{tabular}
\vspace{-2mm}
\label{tab:respectives}
\end{table}

\subsection{Defenses}

The first attempts at obfuscating a design were of the RLL form, an attempt to improve security using XOR/XNOR gates driven by key inputs and internal signals~\cite{roy2008}. RLL can easily achieve high output corruption if enough key gates are inserted. Researchers have then modified the original RLL technique to strengthen its resilience against various attacks. For example, fault analysis-based methods~\cite{rajendran2015} focus on analyzing the susceptibility of the locked circuit to fault attacks, and strong-interference-based LL approaches~\cite{rajendran12} are designed to create interference that is strong enough to confuse or mislead an attacker attempt, ensuring that adversaries cannot easily determine the correct keys by creating complex dependencies between them. However, the emergence of SAT attacks, which iteratively find Distinguishing Input Patterns (DIPs) that rule out wrong keys, has highlighted vulnerabilities in these methods~\cite{subramanyan2015}. A DIP is an input combination that produces different outputs for two distinct key values. Consequently, the DIP provides an unequivocally incorrect key value that can be eliminated. This has led to the exploration of many other alternative LL techniques. 


Figs.~\ref{fig:circuits}(a) and~\ref{fig:circuits}(c) show how the LL techniques are divided into two main categories: SFLT and DFLT, respectively. SFLT uses a single critical signal which corrupts the original circuit. SFLT techniques use a point function in their locking unit, thus forcing attacks to explore an exponential number of queries. The use of a point function makes key recovery increasingly challenging as the SAT problem size grows with each iteration~\cite{yasin2017}. However, the characteristic of having only one critical signal can be exploited for the potential recovery of the original design via a removal attack aimed precisely at this crucial signal~\cite{yasin2020}. In contrast, DFLT techniques introduce an additional critical signal in an attempt to be more resilient against removal attacks. DFLT combines a restore and a perturb unit to secure circuits. The restore unit ensures correct functionality with the proper key, while the perturb unit disrupts operation with an incorrect key. This dual-phase approach enhances security and resists reverse engineering and SAT attacks. Then, even if the restore unit is removed in a DFLT, the original circuit design remains unrecoverable because of the perturb unit, as shown in Fig.~\ref{fig:circuits}(c). Consequently, DFLT provides enhanced protection by introducing an added layer of security, thwarting unauthorized access, and preserving the locked circuit’s integrity.

In the SFLT category, the Anti-SAT~\cite{xie2019} method incorporates a locking unit that executes a complementary function. This unit is structured around two distinct functions, denoted as $g$ and $\bar{g}$, which are ANDed to generate a single critical signal. Subsequently, this critical signal is integrated with the original design by an XOR gate that drives the output of the original design. The Anti-SAT-DTL~\cite{shamsi2019} approach has a diversified tree logic (DTL), where some AND gates in the AND-tree are replaced by OR/NAND/XOR gates. In CASLock~\cite{shakya2019}, the $g$ block is constructed using cascaded AND-OR gates, which serve as its distinctive feature. The effectiveness of CASLock lies in its incorporation of OR gates within the $g$ and $\bar{g}$ blocks. This design attribute enables CASLock to effectively counter-bypass attacks by strategically altering the placement and quantity of AND/OR gates within the $g$ and $\bar{g}$ blocks. SARLock~\cite{yasin2016} introduces a comparator and a masking circuit connected to the original netlist, causing corruption on a specific input pattern.

DFLT techniques, such as stripped functionality logic locking (SFLL), strip some functionality from the original design, corrupting its output corresponding to protected input patterns~\cite{yasin2017, yasin2017_2}. SFLL includes a comparator as a restore unit for particular input vectors, with the ability to utilize a configurable hamming distance (HD) for protecting multiple input vectors~\cite{yasin2017_2}. Note that tenacious and traceless logic locking (TTLock) is equivalent to SFLL when HD is equal to 0~\cite{yasin2017}.

Figs.~\ref{fig:circuits}(b) and~\ref{fig:circuits}(d) exemplify CLL which integrates multiple LL techniques to enhance the security of ICs. Note that RLL is always used as it delivers the important feature of (high) output corruption. By combining RLL with other techniques, CLL endeavors to leverage their respective strengths while mitigating individual weaknesses. This integration strategy aims to fortify security by exploiting the complementary aspects of diverse LL techniques, selecting corruption levels and SAT resilience tailored to optimize the desired trade-off between security and corruptibility.

\subsection{Attacks}

LL has received much attention for being implementable at the front-end stage (i.e., as direct modifications to a netlist) without requiring layout modifications or foundry collaboration (as is the case with split manufacturing). However, its security has been challenged by various emerging attacks that have repeatedly succeeded at exposing secret keys. OG attacks involve comparing a locked design with an activated device, using the activated circuit as an oracle. This allows an adversary to examine the differences between the locked and unlocked designs, enabling them to compare the outputs of the original circuit with those of the locked circuit ~\cite{shen2018, subramanyan2015, rajendran2015}. The most well-known OG attack, the SAT-based attack, iteratively finds DIPs to break LL techniques~\cite{shen2018, subramanyan2015, rajendran2015}. AppSAT reduces the time to breach LL by finding an approximate solution instead of fully solving the SAT problem. At the same time, Double DIP enhances SAT-based attacks by refining key guesses through two decryption steps, exploiting weaknesses in the locking mechanism. Query attacks are another type of SAT-based attack, where each query is applied to the oracle, and the values of primary outputs are obtained~\cite{almeida2023}. Each query $q$ corresponds to a distinct input or a set of inputs furnished by the attacker to the oracle. Upon receipt of a query $q$, the oracle generates a response, denoted as $r$, which embodies the output produced in reaction to the supplied query. This process can be represented as $f(q) = r$, where $f$ denotes the function mapping inputs $q$ to outputs $r$. This information is helpful to an attacker because $r$ implicitly is a function of the key present inside the oracle.

In contrast, OL attacks focus on extracting sensitive information from a locked IC without direct access to an oracle. Adversaries in OL attacks possess only the locked design netlist. Techniques such as machine learning, constant propagation analysis, and resynthesis-based approaches are commonly employed in the OL setting to extract information from the locked IC~\cite{raj2023, alaql2021, almeida2023}. Note that, by definition, OL is a much more difficult setting for the adversary than OG is. The SCOPE attack, which is a prime example of an OL attack, compromises the locked design by leveraging synthesis-based constant propagation. Unlike traditional SAT-based attacks that need an oracle to compare circuits, SCOPE analyzes and simplifies circuits during synthesis. SCOPE identifies and propagates constant values through the circuit, effectively reducing the complexity of the LL and ultimately producing a guess of the correct key with a relatively high degree of certainty.


OG attacks leverage an oracle, amplifying the attacker's capabilities and posing formidable challenges to LL security implementations~\cite{shen2018, subramanyan2015, rajendran2015}. While OG attacks are generally more effective than OL attacks, the assumptions behind this threat model are very favorable to the adversary. Conversely, OL attacks do not enjoy access to the activated IC, and therefore they are generally less effective~\cite{raj2023, alaql2021, almeida2023}. Yet, both attack types have shown effectiveness against specific LL variants~\cite{almeida2023}. We will consider both when introducing our framework RESAA.


\subsection{Cat-and-Mouse Game}
\label{subsection:catandmouse}

In the world of LL, there is an ongoing game between designers working hard to protect ICs and those trying to find vulnerabilities in their defenses. This perpetual cat-and-mouse game is driven by the relentless pursuit of innovation on both sides, where each advancement in defense prompts a corresponding adaptation in attack strategy and vice versa.

It all started with the introduction of LL techniques such as RLL~\cite{roy2008}. However, adversaries quickly responded with a multitude of SAT-based attacks, leveraging powerful algorithms to unlock these protections~\cite{shen2018}. This prompted the development of Anti-SAT~\cite{xie2019} and other countermeasures aimed to thwart such attacks by integrating point functions into the design. However, even as defenders introduced new layers of security, adversaries were quick to devise sophisticated signal probability skew (SPS) attacks~\cite{yasin2020}, exploiting subtle vulnerabilities in the system to bypass defenses.

In response to these emerging threats, researchers engaged in a constant cycle of innovation and adaptation. To bolster their defenses, they implemented novel modifications, such as hard-coding specific input patterns and introducing key-controlled restore units~\cite{yasin2017}. However, each advancement in defense invited a corresponding advance in attack strategy. Adversaries became adept at discerning traces of hard-coded patterns through advanced analysis techniques~\cite{xu2017}, highlighting the escalating arms race between defenders and attackers in LL.

The cycle continued despite the introduction of new proposals like CASLock~\cite{shakya2019} to enhance defenses against SAT-based attacks. Structural attacks~\cite{sengupta2021} eventually found weaknesses in these defenses, demonstrating the vulnerability of even the most advanced security measures. This relentless pursuit of innovation on both sides underscores the never-ending nature of the cat-and-mouse game in LL. Adversaries persist in their quest to exploit weaknesses, while defenders must remain vigilant and adapt to stay one step ahead in the ongoing battle for IC security.
	\section{Proposed Methodology}
\label{sec:methodology}


\begin{figure}[t]
	\centerline{\includegraphics[width=5.5cm]{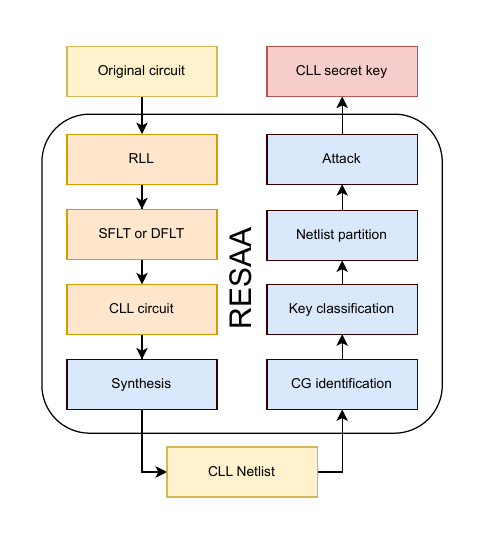}}
    \vspace{-4mm}
	\caption{Overview of the RESAA framework: The left portion shows the pre-processing step to lock and translate the CLL benchmark into mapped Verilog. The CLL netlist is then partitioned and subjected to attacks, revealing the CLL secret key.}
	\vspace*{-4mm}
	\label{fig:classification_0}
\end{figure}

This section describes our attack strategy explicitly aimed at CLL-protected circuits. The scenario begins with a locked design, initially locked with RLL and subsequently secured with a PSLL technique. The attacker -- equipped with reverse engineering capabilities to access the gate-level locked netlist mapped to a commercial library, the functional IC, and EDA tools -- then utilizes RESAA under either the OG or OL scenarios.

We detail our practical classification analysis and the partitioning of the CLL circuit into two distinct netlists facilitated by CG identification. Each netlist includes all the necessary inputs related to a particular LL technique. Our method involves implementing the RESAA attack on these divided netlists, effectively revealing the secret key.

\subsection{Classification and Partition}

Fig.~\ref{fig:classification_0} shows our pre-processing method in the left portion of the RESAA framework, which converts original circuits into locked CLL benchmark files (usually bench files) and then mapping them into Verilog netlists (locked circuits) using a commercial synthesis tool. This step is crucial to make the CLL circuits suitable for analysis during the CG identification and key classification phase. The steps drawn in orange in Fig.~\ref{fig:classification_0} are, conceptually, executed by a defender. They are included here for the sake of completeness.

\begin{figure}[t]
	\centerline{\includegraphics[width=5.5cm]{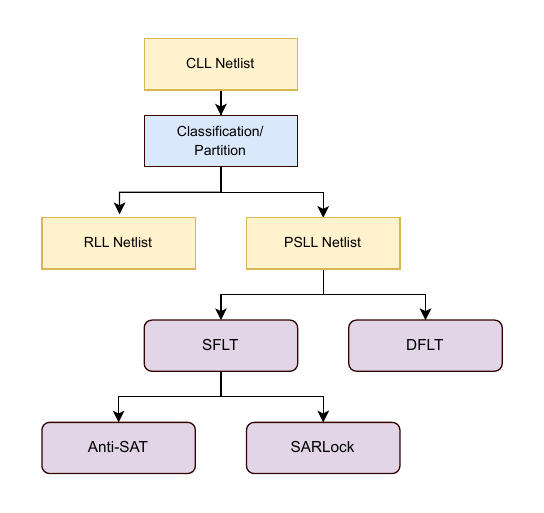}}
	\caption{Classification of techniques employed in a locked netlist.}
	\vspace*{-4mm}
	\label{fig:classification_1}
\end{figure}






RESAA then performs (re)synthesis on the locked design. When doing this, RESAA utilizes a commercial synthesis tool and the same cell library that the defender utilized. However, RESAA introduces one additional restriction: only 2-input gates are allowed during mapping. This restriction simplified the work of RESAA as CGs become easier to identify. Under this restriction, there is always a critical gate where all keys from one technique converge on one input while the other input gathers keys from the other technique.

As seen in Fig.~\ref{fig:icflow}, the CG acts as a common path for all key inputs derived from the RLL and PSLL techniques. By utilizing the internal graph representation provided by the EDA tool, the process was scripted to analyze each key input and group them based on the primary output (PO) they reach. We identify the CG as the first gate where all paths from the key inputs converge before reaching a PO. This gate is crucial for classifying the LL techniques. One input of the CG is solely associated with an RLL key, while the other is associated with PSLL keys.

Our study has identified three distinct behaviors of the key inputs leading to a PO. The first group, which is associated with the use of RLL alone, shows no discernible patterns in the number of reachable outputs. This is predictable, since RLL is supposed to be random. The second group, related to the use of PSLL, is recognized when all key inputs lead to the same number of POs. The third group, a CLL circuit utilizing both RLL and PSLL techniques, emerges when both behaviors are present within a design. 

The relationship between key inputs and POs reached by them is a feature that is leveraged by RESAA to perform netlist partition. Fig.~\ref{fig:classification_1} presents our classification methodology, where RESAA initially partitions the design into two distinct netlists: one containing exclusively RLL key inputs and the other consisting of PSLL key inputs. The PSLL netlist is further categorized into SFLT and DFLT techniques. Within the SFLT category, further classification is performed to distinguish between Anti-SAT and CASLock implementations. It is important to ensure accurate classification of key inputs to have any chance of having a successful attack later on. The following high-level steps outline the classification process:

\begin{figure*}[t]
	\centerline{\includegraphics[width=18.0cm]{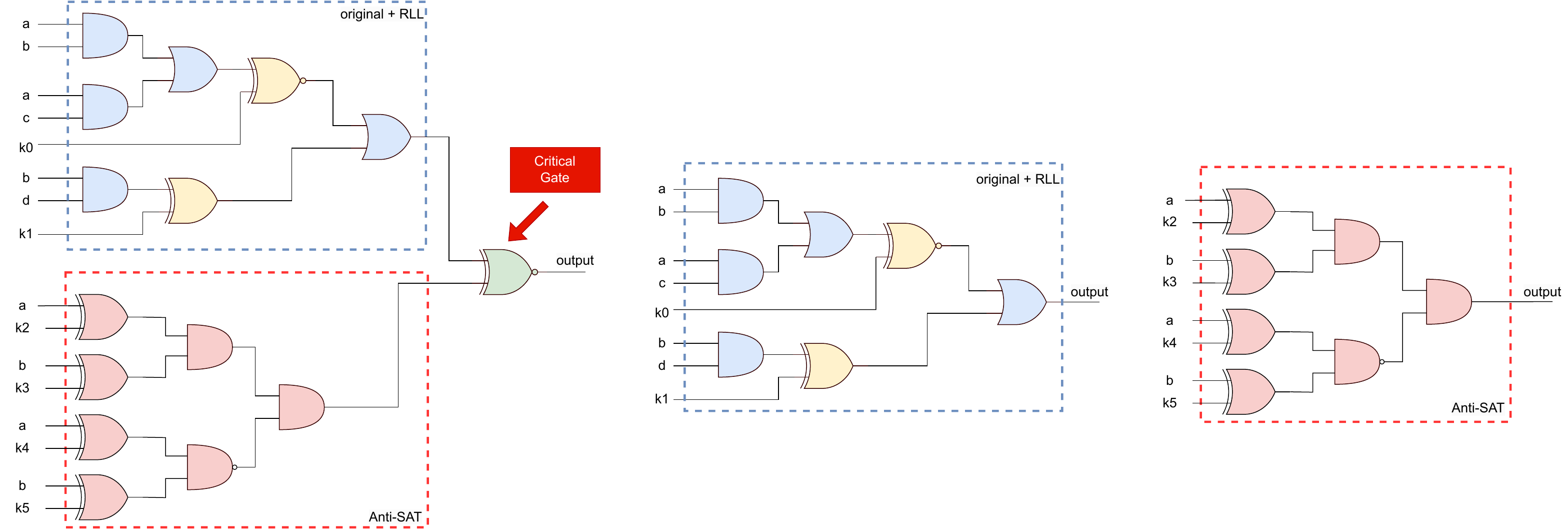}}
	\caption{A circuit locked with RLL and Anti-SAT. The CG identification and partition process by RESAA are highlighted.}
        \vspace*{-2mm}
	\label{fig:detail_circuit}
\end{figure*}

\begin{itemize}
    \item \textbf{Step 1: Generate Key Input Graphs} \\
    Create a directed graph for each key input $k_i$ that shows all paths from $k_i$ to the POs.

    \item \textbf{Step 2: Identify Number of POs for Each Key} \\
    Determine how many POs each key input $k_i$ can reach through the paths.

    \item \textbf{Step 3: Group Key Inputs} \\
    Group key inputs together if they have the same number of reachable POs, suggesting they belong to the same LL technique.

    \item \textbf{Step 4: Identify the CG} \\
    Find the first gate where all paths from the grouped key inputs meet before reaching a PO. This gate is the CG.

    \item \textbf{Step 5: Final Classification} \\
    Check the classification by analyzing the paths through the CG. If a key is misclassified, move to the next gate and try reclassifying. For instance, if an RLL key was wrongly classified as a PSLL key because its paths appeared in both inputs of the CG, the process moves to the next gate. If no key paths match the new gate, the key is correctly reclassified as RLL.
    
\end{itemize}

RESAA takes advantage of the fact that a logic synthesis tool already has, internally, a graph representation of the circuit and its connections that is efficient and can be queried at will. Therefore, Step 1, as previously outlined, requires no effort. With additional scripting, all other steps can be executed within the environment of the synthesis tool itself. This is a key feature of RESAA and contributes to its scalability.

To summarize the inner working of RESAA, Fig.~\ref{fig:detail_circuit} presents an example of a circuit locked by RLL + Anti-SAT. In this example, the CG consolidates all paths from RLL key inputs into one input of the XNOR gate, while all paths from Anti-SAT key inputs are connected to the other input. Following this, RESAA removes this CG, resulting in the generation of two netlists. A first netlist is created by eliminating the CG and directly connecting the RLL partition to the design's PO. Similarly, the second netlist undergoes this process by taking the other input of the CG and connecting it directly to the PO, thereby creating a second netlist containing only the locking/restore.

In a CLL-locked design, the identification of this CG serves as an anchor for splitting the netlist and applying further attacks. In other words, the presence of a CG that consolidates all the various paths from RLL and from the PSLL is a vulnerability. Identifying this critical node yields valuable insights into the structural organization of the locked netlist, enabling the systematic division of the design into two distinct netlists. By identifying the CG, partitioning the netlist into two, and employing appropriate attack(s), an adversary can more easily uncover the secret key than when dealing with the whole CLL-locked circuit. 

\begin{figure}[t]
	\centerline{\includegraphics[width=7cm]{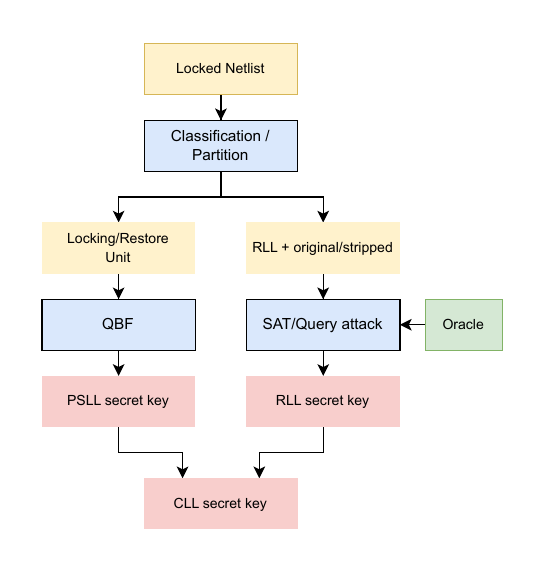}}
    \vspace{-4mm}
	\caption{RESAA attacks under the OG threat model: Netlists are highlighted in yellow, attacks are shown in blue, the oracle is indicated in green, and the resulting secret keys are marked in red.}
	\vspace*{-2mm}
	\label{fig:og_flow}
\end{figure}

Fig.~\ref{fig:classification_0} shows, on the right side, that with the keys now accurately classified into RLL and PSLL categories and the CLL netlist partitioned into two distinct segments, RESAA can proceed to apply specific attacks. In scenarios where the PSLL netlist includes an SFLT, the tool divides the CLL netlist, separating the original netlist with RLL keys and a logic unit from the PSLL technique, as illustrated in Fig.~\ref{fig:circuits}(b). Conversely, if the PSLL consists of a DFLT, RESAA separates the stripped function along with the RLL portion, where the stripped function includes an original circuit along with a perturbation module, as shown in Fig.~\ref{fig:circuits}(c).  

\subsection{Attack Under the OG Threat Model}

Fig.~\ref{fig:og_flow} shows how attacks under the OG threat model can be applied. After classification/partition, we have the netlist consisting of RLL + original/stripped and locking/restore unit. Three attacks are used for this flow: a quantified boolean formula (QBF)-based attack is applied to the locking/restore unit, a SAT-based attack or a query attack is applied to the RLL + original/stripped netlist, where an oracle is used to obtain input/output relationships.


The QBF attack, which will target the locking/restore unit, operates in two main steps. First, it constructs a QBF by combining the conjunctive normal form (CNF) formulas of individual gates, thereby converting the logical structure of the unit into a standardized format suitable for analysis. Secondly, it generates two distinct QBF problems: one for when the output of the locking/restore unit is equal to 0 and another for when this output is 1 for \textbf{all} possible input combinations. 

Subsequently, a QBF solver is employed to these problems. If a solution exists for either of these QBF problems, it signifies that a valid set of key inputs has been identified, effectively revealing the secret key of the partitioned circuit. It is important to highlight that formulating two distinct QBF problems allows for the evaluation of all possible output values. This dual-QBF approach ensures that RESAA can succeed in different scenarios: if we consider the circuit depicted in Fig.~\ref{fig:detail_circuit}, the CG might take the form of an XOR or XNOR, depending on whether the logic formed by the red-colored gates produces a logic 0 or a logic 1, respectively.

When the QBF attack identifies a solution, it confirms the presence of an SFLT within the CLL. In this case, a SAT-based attack is applied to the netlist containing the RLL + original configuration. However, the query attack becomes necessary if the netlist is composed of RLL + stripped. The stripped functionality allows many incorrect keys to produce correct outputs for many inputs, significantly hindering the SAT solver's ability to converge on the proper key. As a result, RESAA can identify the PSLL secret key using the QBF attack and the RLL secret key using the SAT-based attack, thus revealing the entire CLL secret key. In cases where the CLL is composed of RLL + DFLT, QBF alone cannot identify the PSLL secret key, but the query attack can identify many or even the majority of RLL keys.

A combined strategy of SAT-based and query-based attacks was implemented to enhance the effectiveness of RLL key extraction. The process begins by converting the locked netlist, consisting of the RLL + original/stripped configuration, into CNF to facilitate SAT solver analysis. The SAT solver then identifies DIPs that differentiate potential key values. These DIPs are applied to a functional IC, serving as an oracle that provides the correct output for each input. This correct output is then used to refine the CNF formula, progressively eliminating incorrect key candidates. The SAT solver iteratively performs this process until no further DIPs can be found, ensuring that only the correct key values remain.

Subsequently, the flow transitions to a query-based attack phase in cases where no PSLL secret key was found, which uses QBF attack, where specific, strategically crafted queries are directed at the oracle to extract additional insights about the correct key. Each query is analyzed to deduce logical inferences, further refining the CNF formula and pruning the search space. Our combined approach systematically narrows down the possible key values with increased efficiency and precision by targeted queries. Integrating these two attack methodologies ensures a comprehensive and robust evaluation of the (C)LL scheme, ultimately enhancing the likelihood of successful key recovery.

\begin{figure}[t]
	\centerline{\includegraphics[width=7cm]{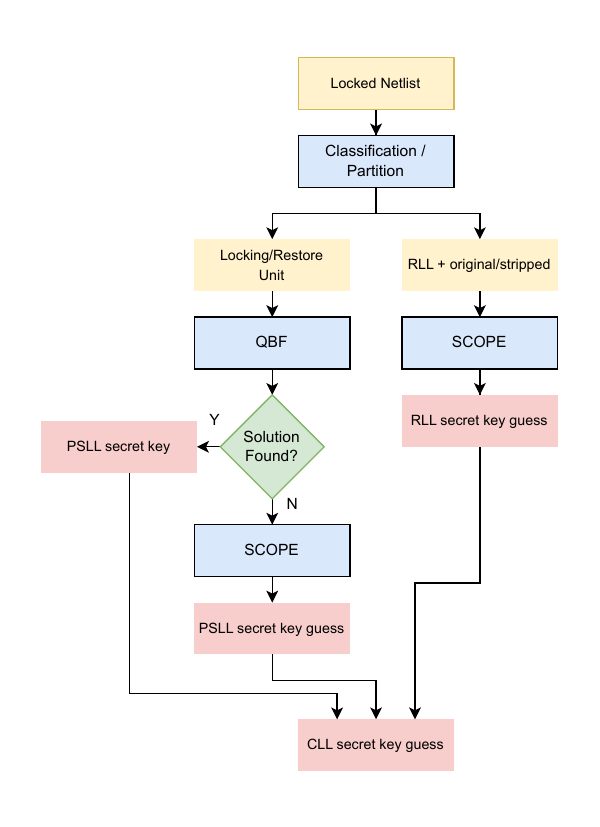}}
    \vspace{-4mm}
	\caption{RESAA steps under the OL threat model: Netlists are highlighted in yellow, attacks are shown in blue, and the resulting secret keys are marked in red.}
	\vspace*{-2mm}
	\label{fig:ol_flow}
\end{figure}

\subsection{Attack Under the OL Threat Model}

Figure~\ref{fig:ol_flow} shows our attack under the OL threat model, which shares its initial steps with the OG threat model and begins with the same CLL netlist as an input. This is followed by a partitioning and classification process that can be adjusted to suit the specific needs of the task. The netlist, now classified and partitioned in a manner identical to the OG model, is subjected to the QBF attack on the locking/restore unit. If this attack is successful, our tool returns the PSLL key, reaffirming the previous classification as SFLT. In the event of an unsuccessful QBF attack, then the OL SCOPE attack is run on this netlist.

When applying SCOPE to the partition containing both RLL + original/stripped and locking/restore unit to find a solution, it is important to note that the SCOPE solution may yield a logical value of 0, 1, or an undetermined solution (`x') for a given key input. If the QBF attack fails to determine the PSLL key, SCOPE is subsequently employed in the locking/restore unit to guess the values of the key inputs.


To bolster our methodology's reliability and ensure the developed tool's correctness, we introduced an additional step involving a Logic Equivalence Checking (LEC) tool, specifically for internal verification. Therefore, this step is not shown in the graphical representation of the attack flow given in Figures~\ref{fig:og_flow} and~\ref{fig:ol_flow}. This integration occurred at two crucial junctures within our flow. First, following partitioning, we employed the LEC tool to compare the RLL with the original design, particularly in scenarios involving SFLT and the preceding netlist locked solely with the RLL we generated. Secondly, after uncovering the secret key for CLL, we utilized the LEC tool to ensure the integrity of our process. This involved comparing the modified CLL netlist, which includes the recovered secret key, against the original netlist. The LEC tool checks for logical equivalence between the two netlists, verifying that the CLL netlist with the secret key produces the same outputs as the original netlist for all possible inputs. We solely utilized this step for verification purposes. An adversary \textbf{does not} possess this capability since, by definition, the adversary does not possess the original circuit.


	\section{Results}
\label{sec:results}

Our methodology utilizes a combination of Perl, Python, and TCL scripts to interface with the commercial logic synthesis tool Cadence Genus~\cite{genus2024}. Synthesis is conducted using a commercial $65$ nm standard cell library. All experiments were performed on a 32-core Intel Xeon processor running at 3.60 GHz with a RAM capacity of 1 TB. Yet, no attacks are multithreaded, and the results presented here can be generalized to a personal computer.

Our study involved the analysis of CLL circuits utilizing a set of ten benchmarks: \textit{c2670}, \textit{c3540}, \textit{c5315}, \textit{c6288}, and \textit{c7552} sourced from the ISCAS'85 benchmark suite~\cite{iscas1985}, along with \textit{b14}, \textit{b15}, \textit{b20}, \textit{b21}, and \textit{b22} from the ITC'99 benchmark suite~\cite{itc99}. 

Table~\ref{tab:report1} presents a comprehensive description of the benchmarks, including the number of inputs (\#in) and outputs (\#out) alongside pertinent metrics such as area ($\mu m^2$), power consumption ($mW$), and delay ($ps$) for each benchmark.

\begin{table}[t]
  \centering
  \footnotesize
  \caption{Details of ISCAS'85 and ITC'99 circuits.}
  \vspace{-2mm}
  \begin{tabular}{|@{\hskip2pt}c@{\hskip2pt}|c@{\hskip2pt}c@{\hskip2pt}c@{\hskip2pt}c@{\hskip2pt}c|c@{\hskip2pt}|c@{\hskip2pt}c@{\hskip2pt}c|}
    \hline
    \multirow{2}{*}{Circuit} & \multicolumn{5}{c|}{\multirow{1}{*}{Original Netlist}} &  \multirow{2}{*}{$\#p$} & \multicolumn{3}{c|}{RLL Locked Netlist} \\ 
    \cline{2-6} \cline{8-10}
    & \#in & \#out & area & power & delay & & area & power & delay \\ 
    \hline \hline
c2670	&	157	&	64	&	1046	&	3.36	&	1264	&	64	&	1424	&	5.22	&	1742	\\
c3540	&	50	&	22	&	1518	&	6.58	&	1977	&	32	&	1655	&	7.48	&	2091	\\
c5315	&	178	&	123	&	2460	&	9.9	&	1864	&	64	&	2864	&	1.21	&	1982	\\
c6288	&	32	&	32	&	3133	&	8.48	&	4621	&	32	&	3303	&	9.09	&	5160	\\
c7552	&	206	&	105	&	2702	&	1.32	&	1663	&	64	&	3209	&	1.61	&	2015	\\    
b14	    &  275	&	245	&	8326	&	3.59	&	4882	&	128	&	8872	&	4.34	&	4864	\\
b15	    &  485	&	449	&	12416	&	3.11	&	4809	&	128	&	12938	&	3.78	&	5188	\\
b20	    &  522	&	512	&	17210	&	8.95	&	5536	&	128	&	17731	&	9.93	&	5511	\\
b21	    &  522	&	512	&	17685	&	9.24	&	5075	&	128	&	18248	&	9.94	&	5061	\\
b22	    &  767	&	757	&	26416	&	1.33	&	5358	&	128	&	26941	&	1.39	&	5282	\\
    \hline
  \end{tabular}
  \label{tab:report1}
\end{table}

Moreover, Table~\ref{tab:report1} also presents the area, power, and delay values after the benchmarks were locked by the RLL technique using the Neos tool~\cite{neos}. The methodology for determining the number of RLL key inputs ($\#p$) was guided by considering both the number of inputs and the overhead associated with the LL technique. In general, 128 keys were used in ITC'99 circuits, and 32 or 64 keys were used in ISCAS'85.

Analysis of the data presented in Table~\ref{tab:report1} reveals a consistent trend across all benchmarks subjected to an RLL scheme. There is a modest increase in area and power consumption, averaging approximately 4.5\%, alongside an average delay increment of about 10.5\%. This trend aligns with the characteristics of the RLL technique (insertion of XOR/XNOR gates), which minimally affects area and power consumption. Nevertheless, incorporating additional logic gates along the critical path contributes to a slight rise in delay.

After the RLL locking phase, the second locking technique was introduced by utilizing one out of five distinct PSLL techniques, namely Anti-SAT, Anti-SAT-DTL, CASLock, and SARLock as SFLTs and TTLock as a DFLT. The implementation of Anti-SAT, Anti-SAT-DTL, and TTLock was facilitated using the Neos tool~\cite{neos}. Meanwhile, SARLock was implemented using a Python script developed by P. Subramanyan, and CASLock was implemented using a Perl script by L. Aksoy.

Table~\ref{tab:report2} presents the total number of key inputs ($\#k$) alongside associated metrics, including area, power, and delay across all considered CLL benchmarks. The area exhibited an average increase of 4.50\% compared to the original version. An average increase of 10\% in the CLL of SFLT and around 11.6\% when composed with DFLT. Notably, power consumption remains relatively stable compared to benchmarks locked with only RLL. Moreover, the observed delay overheads are approximately 5\% for SFLT and around 8\% for RLL+TTLock configurations when compared to the netlist locked with only RLL.

\begin{table*}[t]
  \centering
  \footnotesize
  \caption{Details of ISCAS'85 and ITC'99 circuits locked with CLL.}
  \vspace{-2mm}
  \begin{tabular}{|@{\hskip2pt}c@{\hskip2pt}|c@{\hskip2pt}|c@{\hskip2pt}c@{\hskip2pt}c|c@{\hskip2pt}c@{\hskip2pt}c|c@{\hskip2pt}c@{\hskip2pt}c|c@{\hskip2pt}c@{\hskip2pt}c|c@{\hskip2pt}c@{\hskip2pt}c|}
    \hline
    \multirow{3}{*}{Circuit} & \multirow{3}{*}{$\#k$} & \multicolumn{15}{c|}{CLL Locked Netlist} \\ 
    \cline{3-17}
    & & \multicolumn{3}{c|}{RLL+Anti-SAT} & \multicolumn{3}{c|}{RLL+Anti-SAT-DTL} & \multicolumn{3}{c|}{RLL+CASLock} & \multicolumn{3}{c|}{RLL+SARLock} & \multicolumn{3}{c|}{RLL+TTLock} \\ 
    \cline{3-17}
    & & area & power & delay & area & power & delay & area & power & delay & area & power & delay & area & power & delay \\
    \hline \hline
c2670	&	128	&	1792	&	0.63	&	1727	&	1789	&	0.64	&	1836	&	1793	&	0.65	&	1754	&	1842	&	0.65	&	1842	&	1748	&	0.64	&	1788	\\
c3540	&	64	&	1813	&	0.80	&	2083	&	1826	&	0.81	&	2052	&	1846	&	0.83	&	2042	&	1886	&	0.82	&	2087	&	1840	&	0.81	&	2096	\\
c5315	&	128	&	3226	&	1.33	&	1954	&	3213	&	1.32	&	2070	&	3244	&	1.35	&	1980	&	3280	&	1.34	&	2081	&	3198	&	1.32	&	2005	\\
c6288	&	64	&	3471	&	9.15	&	5233	&	3446	&	9.13	&	5119	&	3477	&	9.15	&	5167	&	3524	&	9.16	&	5250	&	3508	&	9.23	&	5106	\\
c7552	&	128	&	3861	&	1.84	&	2109	&	3577	&	1.73	&	2011	&	3378	&	1.68	&	2004	&	3623	&	1.75	&	1993	&	3542	&	1.77	&	1972	\\
b14	&	256	&	9557	&	4.58	&	4600	&	9526	&	4.56	&	4640	&	9637	&	4.62	&	4862	&	9820	&	4.63	&	4893	&	10052	&	4.65	&	4949	\\
b15	&	256	&	13608	&	4.00	&	5148	&	13591	&	4.00	&	5144	&	13695	&	4.04	&	5020	&	13897	&	4.07	&	5036	&	14476	&	4.43	&	5067	\\
b20	&	256	&	18379	&	10.14	&	5497	&	18379	&	10.12	&	5470	&	18468	&	10.20	&	5609	&	18641	&	10.22	&	5488	&	20023	&	10.91	&	5728	\\
b21	&	256	&	18904	&	10.17	&	4922	&	18893	&	10.17	&	4954	&	18968	&	10.21	&	5030	&	19167	&	10.23	&	5045	&	20069	&	10.98	&	5127	\\
b22	&	256	&	27559	&	14.02	&	5376	&	27578	&	14.03	&	5367	&	27703	&	14.19	&	5317	&	27803	&	14.10	&	5290	&	30007	&	14.51	&	5421	\\
    \hline
  \end{tabular}
  \label{tab:report2}
\end{table*}

Initially, we performed several attacks documented in the existing literature to compare against our developed methodology, expecting all CLL circuits to withstand SAT-based attacks and their variants. All benchmarks locked in the CLL scheme were submitted to four attacks: the SAT-based attack developed by~\cite{subramanyan2015}, Double-DIP (DP) implemented by~\cite{shen2017}, query attack QATT by~\cite{qatt2023}, and AppSAT, an evolution of the SAT attack, implemented by~\cite{vijayakumar2017}.

Table~\ref{tab:allcll} shows the runtime to find a solution, with ``out-of-time'' (OoT) indicating instances where no solution could be found within the allowed 48-hour time limit. As observed from Table~\ref{tab:allcll}, the SAT-based and DP attacks exhibit low efficiency in deciphering key inputs, as expected. The SAT-based attack only found a solution for one single RLL+TTLock case in the \textit{c5315} circuit. The AppSAT attack showed promising results for small circuits but demanded significant execution time compared to other attacks. While the approach demonstrated near 100\% efficiency for ISCAS'85 circuits, it failed to solve any cases for ITC'99 circuits. This limitation arises due to the exponential increase in complexity with circuit size, making it computationally prohibitive to determine the correct key values that satisfy the SAT solver.

Lastly, our query attack from~\cite{qatt2023}, QATT, displayed varying execution times and degrees of success in deciphering key inputs, as shown in Table~\ref{tab:allcll}. The number of proven key inputs $prv$ discovered ranged just over 41\%, with execution times spanning from 8 to 2645 seconds.

It is important to note that these attacks are available from the literature, and when designed, initially targeted circuits that are locked with a specific LL technique. Interestingly, according to Table~\ref{tab:allcll}, these attacks did not obtain significant results in uncovering key inputs in experiments carried out with CLL circuits. The AppSAT attack, despite its potential to find solutions quickly, presents significant challenges. In other words, available attacks often prove highly inaccurate and ultimately ineffective, mainly when dealing with intricate CLL structures. In this context, we emphasize that a tool like RESAA is invaluable. Next, we present the results of our OG and OL attack strategies. 

\subsection{Results of RESAA under the OG Threat Model}

Under the OG threat model, two well-known attacks were considered. Specifically, the QBF attack outlined in~\cite{aksoy2024} was employed to decipher PSLL keys, along with a SAT-based attack from~\cite{sat2018} and a query attack from~\cite{aksoy2024} to manage RLL key inputs.

\begin{figure*}[t!]
    \centering
    \begin{subfigure}[t]{0.5\textwidth}
        \centering
        \includegraphics[width=0.97\linewidth]{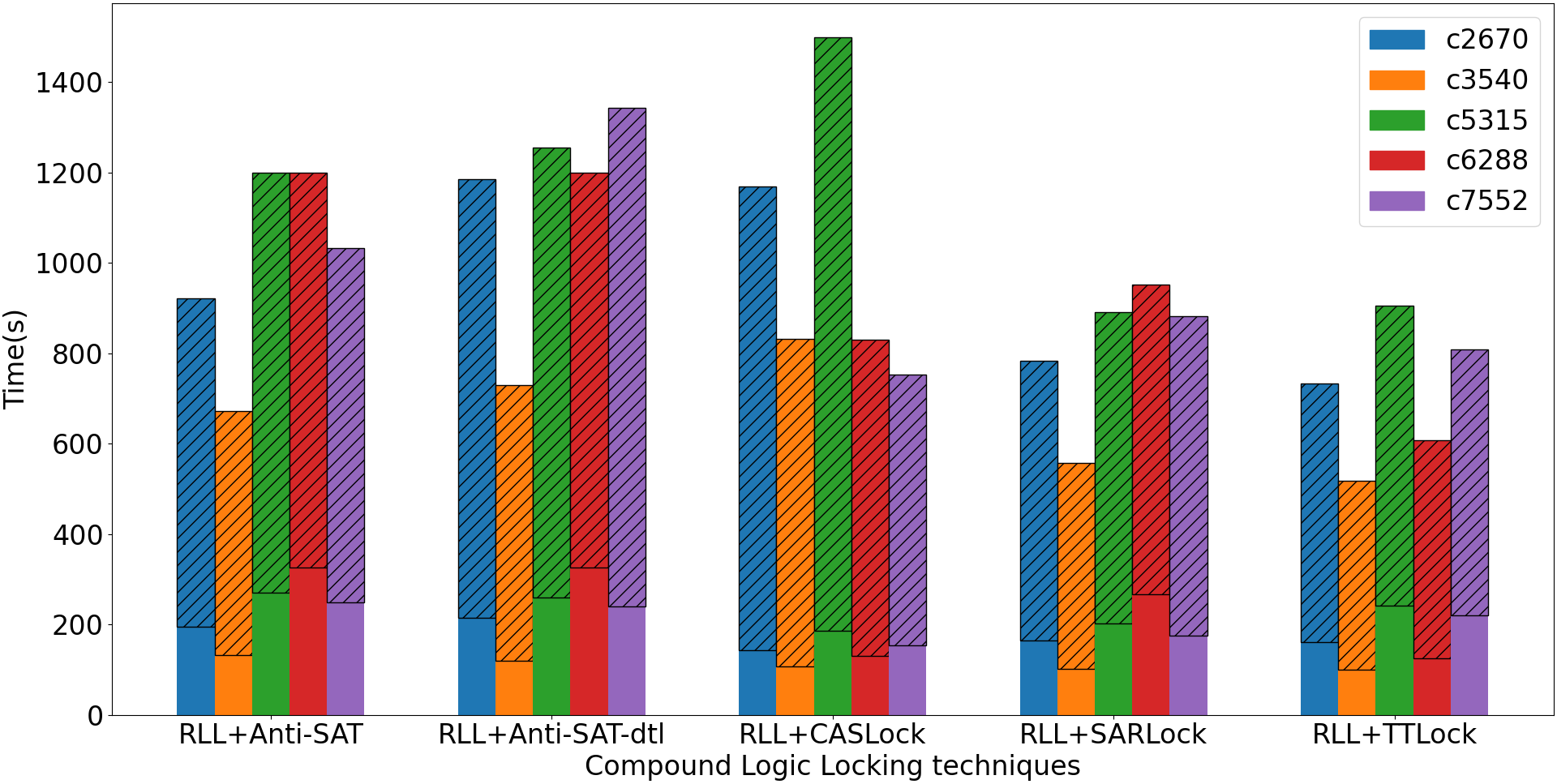}
        \caption{ISCAS’85 benchmark}
    \end{subfigure}%
    ~ 
    \begin{subfigure}[t]{0.5\textwidth}
        \centering
        \includegraphics[width=0.97\linewidth]{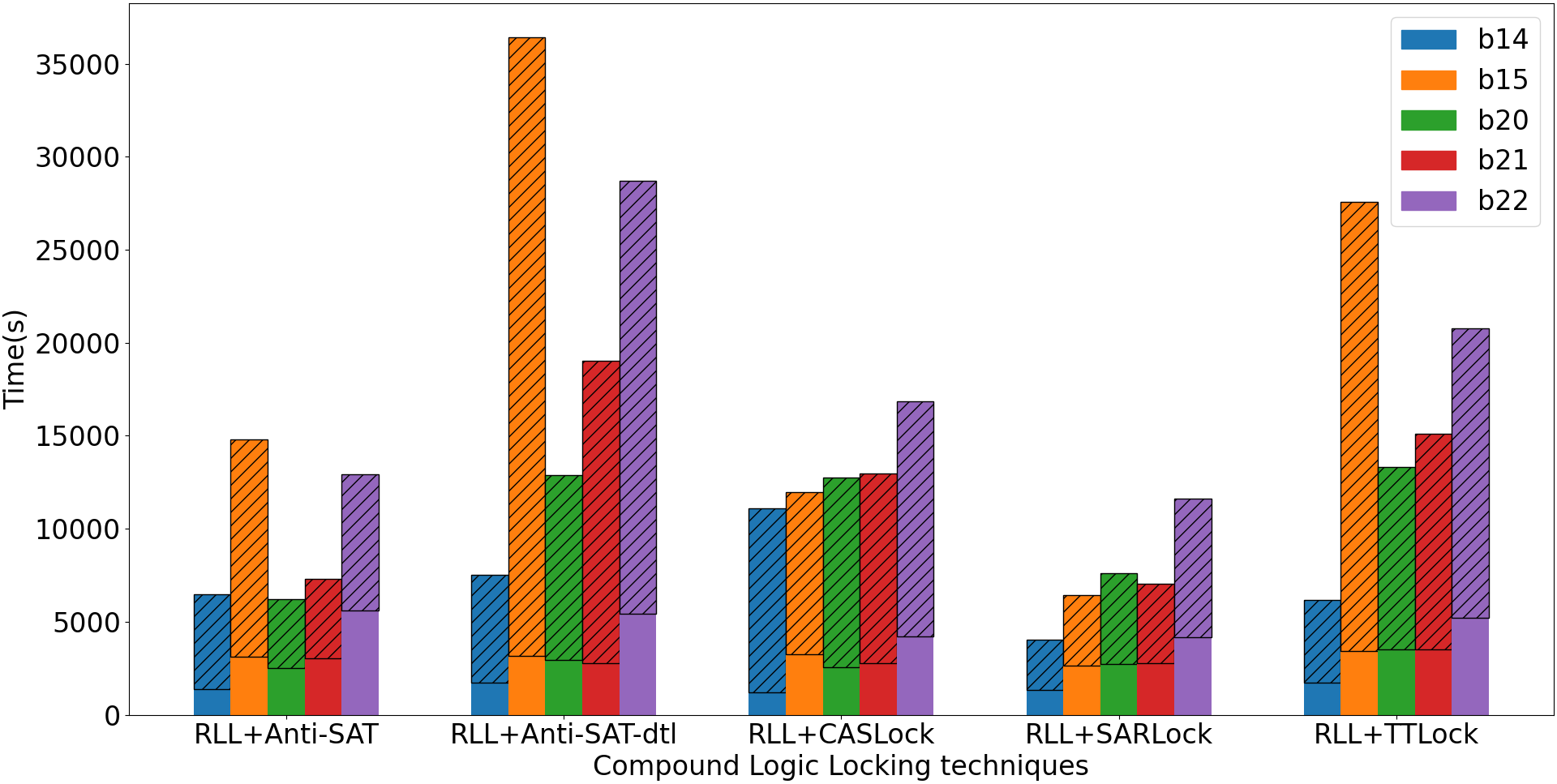}
        \caption{ITC'99 benchmark}
    \end{subfigure}
    \caption{Average classification and execution times (seconds) for attacking ISCAS'85 and ITC'99 benchmarks in the CLL scheme. Bottom: Classification and partition time. Hatched: Attack time. Combined: Total execution time.}
    \label{fig:class}
\end{figure*}

Fig.~\ref{fig:class} presents the classification and execution times. In this context, ``classification time'' refers to the duration required for categorizing the LL technique utilized in the CLL design, depicted in the lower section of the graph. Conversely, ``attack time'' is indicated by the hatched portion of the graph, while `execution time' represents the total time, including both classification/partition time and the subsequent attack on each circuit, depicted by both sections in the graph.

Upon observation, it is evident that when a CLL design is classified as RLL + SFLT, a solution emerges during the QBF attack of the locking/restore unit. Subsequently, the netlist consisting of RLL + original becomes vulnerable to the SAT-based attack. RESAA successfully deciphered all key inputs for CLL circuits with SFLT. In cases where no solution is found following the QBF attack, the second netlist is mandatorily classified as RLL + stripped. In this scenario, a query attack is applied to the RLL + stripped netlist, resulting in guessed RLL key inputs.

In each case, the complete set of key inputs was successfully exposed by implementing the partitioning approach, achieving 100\% discovery when CLL included both RLL and SFLT. However, when a DFLT was introduced as a second technique, initial attempts to uncover PSLL key inputs were unsuccessful. It was only after employing a query attack that some of the RLL keys were eventually disclosed. A time limit of 1-hour was set for this query attack, as further execution did not yield improved results despite prolonging the runtime. We do hypothesize that this could be improved by changing the query strategy.

Our classification and partitioning step involves processing a CLL netlist as input, where a timing analysis\footnote{Timing analysis here is meant by the STA performed by the logic synthesis tool. By performing STA, inherently a graph is built that can be used to query whether an input $i$ has a path to an output $o$.} is conducted to distinguish between RLL and PSLL key inputs, as illustrated in Fig.~\ref{fig:og_flow}. The size and complexity of the CLL design directly influence the duration of both the classification and execution phases. For instance, the maximum execution time for ISCAS'85 benchmarks was approximately \textit{1400} seconds, whereas it reached around \textit{36000} seconds for the larger ITC'99 benchmarks. The classification time typically accounts for less than \textit{1/3} of the total execution time.

The validation process, using the LEC tool, was crucial in cases where all the keys were revealed to validate RESAA. In each instance, the netlist composed of the RLL + original portion was equivalent to the netlist previously locked with RLL. Additionally, the CLL netlist, with the added secret key inputs, was confirmed to coincide in functionality with the original design, thereby certifying the accuracy of RESAA in the partitioning processes. This result confirms a high level of confidence in RESAA's outcomes.

\begin{table*}[t]
  \centering
  \footnotesize
  \caption{Details of existing attacks in ISCAS'85 
 and  ITC'99 circuits locked using a CLL scheme.}
  \vspace{-2mm}
  \begin{tabular}{|@{\hskip2.0pt}c@{\hskip2.0pt}|c@{\hskip2.0pt}c@{\hskip2.0pt}c@{\hskip2.0pt}c@{\hskip2.0pt}c@{\hskip2.0pt}|c@{\hskip2.0pt}c@{\hskip2.0pt}c@{\hskip2.0pt}c@{\hskip2.0pt}c@{\hskip2.0pt}|c@{\hskip2.0pt}c@{\hskip2.0pt}c@{\hskip2.0pt}c@{\hskip2.0pt}c@{\hskip2.0pt}|c@{\hskip2.0pt}c@{\hskip2.0pt}c@{\hskip2.0pt}c@{\hskip2.0pt}c@{\hskip2.0pt}|c@{\hskip2.0pt}c@{\hskip2.0pt}c@{\hskip2.0pt}c@{\hskip2.0pt}c@{\hskip2.0pt}|}
    \hline
    \multirow{4}{*}{Circuit} & \multicolumn{25}{c|}{Locked Netlist} \\ 
    \cline{2-26} 
    & \multicolumn{5}{c|}{RLL+Anti-SAT} & \multicolumn{5}{c|}{RLL+Anti-SAT-DTL} & \multicolumn{5}{c|}{RLL+CASLock} & \multicolumn{5}{c|}{RLL+SARLock} & \multicolumn{5}{c|}{RLL+TTLock} \\ 
    \cline{2-26}
    & sat & appsat & dp & \multicolumn{2}{c|}{qatt} & sat & appsat & dp & \multicolumn{2}{c|}{qatt} & sat & appsat & dp & \multicolumn{2}{c|}{qatt} & sat & appsat & dp & \multicolumn{2}{c|}{qatt} & sat & appsat & dp & \multicolumn{2}{c|}{qatt}\\
    \cline{2-26}
    & time & time & time & prv & time & time & time & time & prv & time & time & time & time & prv & time & time & time & time & prv & time & time & time & time & prv & time\\ 
    \hline \hline
    c2670 & OoT & 998  & OoT & 52 & 48 & OoT & 170  & OoT & 50 & 45 & OoT & 238  & OoT & 50 & 25.4 & OoT & 114  & OoT & 55 & 59 & OoT & 1055  & OoT & 48 & 48 \\
    c3540 & OoT & 766  & OoT & 30 & 20 & OoT & 203  & OoT & 32 & 19 & OoT & 66  & OoT & 30 & 8 & OoT & 91  & 5 & 31 & 22 & 48656 & 4499 & 2 & 31 & 19 \\
    c5315 & OoT & 239  & OoT & 62 & 50 & OoT & 9949  & OoT & 62 & 52 & OoT & 131  & OoT & 62 & 42 & OoT & 64  & OoT & 62 & 72 & OoT & 2768  & OoT & 62 & 60 \\
    c6288 & OoT & OoT  & OoT & 32 & 77 & OoT & 58470  & OoT & 32 & 69 & OoT & 3936  & OoT & 32 & 90 & OoT & 935  & OoT & 32 & 119 & OoT & 6270  & OoT & 32 & 93 \\
    c7552 & OoT & 172  & OoT & 55 & 96 & OoT & 543  & OoT & 55 & 105 & OoT & 279  & OoT & 55 & 59 & OoT & 247  & OoT & 55 & 108 & OoT & 51  & OoT & 55 & 79 \\
    b14 & OoT & OoT  & OoT & 109 & 1822 & OoT & OoT  & OoT & 112 & 1383 & OoT & OoT  & OoT & 106 & 2082 & OoT & OoT  & OoT & 109 & 1335 & OoT & OoT  & OoT & 111 & 1783 \\
    b15 & OoT & OoT  & OoT & 98 & 466 & OoT & OoT  & OoT & 97 & 635 & OoT & OoT  & OoT & 96 & 836 & OoT & OoT  & OoT & 80 & 513 & OoT & OoT  & OoT & 87 & 671 \\
    b20 & OoT & OoT  & OoT & 115 & 1562 & OoT & OoT  & OoT & 115 & 1647 & OoT & OoT  & OoT & 109 & 2645 & OoT & OoT  & OoT & 115 & 1645 & OoT & OoT  & OoT & 113 & 2187 \\
    b21 & OoT & OoT  & OoT & 113 & 1119 & OoT & OoT  & OoT & 114 & 1337 & OoT & OoT  & OoT & 113 & 1935 & OoT & OoT  & OoT & 110 & 1385 & OoT & OoT  & OoT & 109 & 1738 \\
    b22 & OoT & OoT  & OoT & 113 & 1034 & OoT & OoT  & OoT & 113 & 1214 & OoT & OoT  & OoT & 108 & 1489 & OoT & OoT  & OoT & 105 & 938 & OoT & OoT  & OoT & 111 & 1342 \\    
    \hline
  \end{tabular}
  \label{tab:allcll}
\end{table*}

\subsection{Results of RESAA under the OL Threat Model}

The attack strategy under the OL threat model is more restrictive than the OG model because it does not use an oracle. The SCOPE attack, described in~\cite{scope2020}, and the QBF method from~\cite{aksoy2024} were employed. Specifically, in the OL threat model, the SCOPE attack was exclusively used to estimate key inputs for both RLL and PSLL. This approach was necessary when finding a solution in the netlist composed of the locking/restore unit was not feasible, as illustrated in Fig.~\ref{fig:ol_flow}. 

Table~\ref{tab:olattack} presents the results of the SCOPE attack conducted on both the entire CLL design and the partitioned netlist generated by RESAA. In this table, \textit{cdk} and \textit{dk} represent the count of correctly deciphered key inputs and the total deciphered key inputs, respectively, while \textit{time} indicates the overall time taken for the attack.

From Table~\ref{tab:olattack}, it is clear that the SCOPE attack does not succeed in breaking the CLL designs locked by Anti-SAT, Anti-SAT-DTL, and CASLock. However, using RESAA, many key inputs associated with the RLL + original netlists, and even the entire RLL key, can be uncovered. That is exemplified by the \textit{c3540} circuit locked by Anti-SAT-DTL. It is worth noting that the SCOPE attack can reveal over 48\% of key bits using the partition netlists locked with any LL technique. The SCOPE attack on RLL + SARLock circuits resulted in a \textit{dk} rate of 50\%, with an average success rate of 56\% in \textit{dk}. In contrast, RESAA demonstrated a higher \textit{dk}, with cases such as \textit{c5315} achieving a \textit{cdk} rate of 78\%. Furthermore, when a circuit is locked using DFLT, as in RLL + TTLock, the SCOPE attack showed a higher \textit{dk}, but the \textit{cdk} remained lower compared to that achieved by RESAA. In other words, SCOPE alone makes more incorrect guesses than SCOPE post RESAA partitioning.

The runtime of the SCOPE attack, whether executed on the entire design or the partitioned design, significantly depends on the number of gates and keys present in the locked design. Consequently, the runtime for the netlist generated by RESAA after the partition step is notably shorter, as it contains only a fraction of the key inputs and gates. This reduction in complexity results in a smaller runtime, making the attack more efficient. Moreover, the runtime is typically smaller for the partitioned netlist due to the reduced number of gates, as it represents only a portion of the entire design.

The RESAA results for the ITC'99 \textit{b22} circuit locked with RLL + CASLock stand out, achieving 190 \textit{cdk} out of 205 \textit{dk}, a 92.6\% accuracy. These results are very encouraging, given the size and complexity of the circuit paired with the restrictive OL setting. Success in this case demonstrates RESAA's ability to analyze and navigate complex CLL designs effectively. This highlights RESAA's superior performance in accurately recovering keys, even in more complex CLL schemes.


\begin{table*}[t]
	\centering
	\footnotesize
	\caption{Results of OL Attacks on the locked ISCAS'85 and ITC'99  circuits.}
    \vspace{-2mm}
	\begin{tabular}{|@{\hskip2pt}c@{\hskip2pt}|c@{\hskip2pt}c@{\hskip2pt}|c@{\hskip2pt}c@{\hskip2pt}|c@{\hskip2pt}c@{\hskip2pt}|c@{\hskip2pt}c@{\hskip2pt}|c@{\hskip2pt}c@{\hskip2pt}|c@{\hskip2pt}c@{\hskip2pt}|c@{\hskip2pt}c@{\hskip2pt}|c@{\hskip2pt}c@{\hskip2pt}|c@{\hskip2pt}c@{\hskip2pt}|c@{\hskip2pt}c@{\hskip2pt}|}
		\hline
		\multirow{3}{*} {Circuit}  & \multicolumn{4}{c|}{RLL+Anti-SAT} & \multicolumn{4}{c|}{RLL+Anti-SAT-DTL} & \multicolumn{4}{c|}{RLL+CASLock} & \multicolumn{4}{c|}{RLL+SARLock} & \multicolumn{4}{c|}{RLL+TTLock}\\
		\cline{2-21}
		& \multicolumn{2}{c|}{SCOPE} & \multicolumn{2}{@{\hskip2pt}c|}{RESAA} & \multicolumn{2}{c|}{SCOPE} & \multicolumn{2}{c|}{RESAA} & \multicolumn{2}{c|}{SCOPE} & \multicolumn{2}{c|}{RESAA} & \multicolumn{2}{c|}{SCOPE} & \multicolumn{2}{c|}{RESAA} & \multicolumn{2}{c|}{SCOPE} & \multicolumn{2}{c|}{RESAA}\\
		\cline{2-21}
		& cdk/dk & time & cdk/dk & time & cdk/dk & time & cdk/dk & time & cdk/dk & time & cdk/dk & time & cdk/dk & time & cdk/dk & time & cdk/dk & time & cdk/dk & time\\
		\hline \hline
		c2670     & 0/0 & 14  & 73/104 & 4  & 0/0 & 13  & 80/104 & 4  & 0/0 & 9  & 84/104 & 4  & 32/64 & 8   & 97/105 & 4  & 16/106 & 10   & 68/105 & 9 \\
  	    c3540     & 0/0 & 6  & 37/43 & 2  & 0/0 & 6  & 40/64 & 2  & 0/0 & 4   & 
        40/45 & 2  & 15/32 & 4   & 41/43 & 2  & 8/42 & 7   & 32/45 & 4 \\
		c5315     & 0/0 & 15  & 77/107 & 5  & 0/0 & 16  & 78/107 & 5  & 0/0 & 12  & 81/107 & 5  & 32/64 & 10   & 100/107 & 5  & 30/107 & 12   & 69/109 & 11 \\
		c6288     & 0/0 & 7  & 41/56 & 3  & 0/0 & 7  & 42/56 & 2  & 0/0 & 5   & 42/56 & 3  & 17/32 & 5   & 49/56 & 3  & 4/53 & 6   & 40/56 & 5 \\
		c7552     & 0/0 & 17  & 78/108 & 5  & 0/0 & 17  & 79/105 & 5  & 0/0 & 8   & 81/108 &  4 & 40/64 & 11   & 98/107 & 6  & 28/107 & 13   & 82/109 & 12 \\
		b14       & 0/0 & 91  & 160/210 & 67  & 0/0 & 91  & 168/215 & 69  & 0/0 & 67   & 180/213 & 44  & 68/128 & 71   & 168/200 & 46  & 36/203 & 87   & 38/72 & 58 \\
		b15       & 0/0 & 117  & 166/210 & 87  & 0/0 & 120  & 190/214 & 90  & 0/0 & 88   & 180/210 & 58  & 72/128 & 89   & 186/214 & 59  & 49/202 & 109   & 58/82 & 72 \\
		b20       & 0/0 & 155  & 172/203 & 115  & 0/0 & 156  & 182/200 & 117  & 0/0 & 116   & 191/211 & 77  & 67/128 & 118   & 188/209 & 77  & 44/201 & 136   & 62/86 & 93 \\
		b21       & 0/0 & 159  & 183/210 & 119  & 0/0 & 164  & 185/212 & 122  & 0/0 & 119   & 173/213 & 78  & 77/128 & 121   & 178/200 & 82  & 52/196 & 142   & 60/70 & 97 \\
		b22       & 0/0 & 231  & 180/209 & 172  & 0/0 & 233  & 185/212 & 175  & 0/0 & 177   & 190/205 & 116  & 86/128 & 175   & 189/199 & 116  & 48/192 & 193   & 56/70 & 139 \\  
		\hline
	\end{tabular}
	\label{tab:olattack}
\end{table*}


	\section{Conclusions}
\label{sec:conclusions}
This paper introduced the RESAA framework, a novel and comprehensive approach to evaluating and attacking CLL designs. The semiconductor industry's shift towards a fabless model has necessitated advanced security measures to combat emerging threats such as piracy and counterfeiting. CLL, which integrates multiple LL techniques, has been proposed by researchers as a robust solution to these security challenges. However, the security of CLL itself has not been extensively analyzed until now.

Our RESAA framework addresses this gap by systematically classifying locked designs, identifying critical gates, and executing attacks to uncover secret keys. Unlike previous methods, RESAA is agnostic to specific LL techniques, making it a versatile tool for evaluating a wide range of CLL implementations. Through our detailed methodology, which includes classification, partitioning, and applying both OG and OL attack strategies, we demonstrated the framework's ability to expose vulnerabilities in CLL-protected circuits.

Experimental results using ISCAS’85 and ITC’99 benchmark suites highlighted the efficacy of RESAA. The framework successfully identified critical points within the CLL designs, enabling attacks that revealed the secret keys. Our findings underscore the necessity of careful evaluation and selection of LL techniques to ensure the security of IC. The results indicated that even advanced CLL strategies are susceptible to the RESAA framework's targeted attacks.


 
	
	\bibliographystyle{IEEEtran}
	\bibliography{references}
\end{document}